\documentclass[fontsize=9pt,DIV=calc,a4paper,twocolumn]{scrartcl}

\usepackage[T1]{fontenc}
\usepackage[table]{xcolor}
\usepackage[format=plain]{caption}
\usepackage{microtype,amsmath,amssymb,mathtools,physics,textcomp,booktabs,siunitx,multirow,array,xtab,enumitem,cuted,upgreek,tikz}
\usepackage[a4paper,left=15mm,right=15mm,top=15mm,bottom=20mm]{geometry}
\usepackage[auth-lg,affil-it]{authblk}
\usepackage[sort&compress,numbers]{natbib}
\usepackage[hidelinks]{hyperref}
\setlist{noitemsep,topsep=0pt,parsep=0pt,partopsep=0pt,leftmargin=*}

\setkomafont{disposition}{\normalsize\mdseries\bfseries}
\setkomafont{subsection}{\normalsize\mdseries\itshape}

\def\a{-1}
\def\b{0}
\newcommand{\cc}[1]{%
   \pgfmathparse{(min(\b,max(\a,#1))-(\a))/((\b)-(\a))*30}\xdef\darkness{\pgfmathresult}%
   \cellcolor{black!\darkness}#1%
}

\newcommand*\diff{\mathop{}\!\mathrm{d}}

\def\tabscale{0.82}

\title{Experimental parameter uncertainty in PEM fuel cell modeling\\Part II: Sensitivity analysis and importance ranking}

\begin{document}

\author{Roman Vetter}
\author{J\"{u}rgen O.~Schumacher}
\affil{Institute of Computational Physics (ICP),\\Zurich University of Applied Sciences (ZHAW),\\Wildbachstrasse 21, CH-8401 Winterthur, Switzerland}

\twocolumn[
\begin{@twocolumnfalse}
\maketitle
\begin{abstract}
Numerical modeling of proton exchange membrane fuel cells is at the verge of becoming predictive. A crucial requisite for this, though, is that material properties of the membrane-electrode assembly and their functional dependence on the conditions of operation are known with high precision. In this bipartite paper series we determine the most critical transport parameters for which accurate experimental characterization is required in order to enable the simulation of fuel cell operation with sufficient confidence from small to large current densities. In Part II, we employ the two-phase model developed in Part I to carry out extensive forward uncertainty propagation analyses. These include the study of \emph{local} parameter sensitivity in the vicinity of a baseline parameter set, and a \emph{global} sensitivity analysis in which a broad range of operating conditions and material properties is covered. A comprehensive ranking list of model parameters is presented, sorted by impact on predicted fuel cell properties such as the current-voltage characteristics and water balance. The top five in this list are, in this order: The membrane hydration isotherm, the electro-osmotic drag coefficient, the membrane thickness, the water diffusivity in the ionomer and its ionic conductivity.
\end{abstract}
\vskip\baselineskip
\end{@twocolumnfalse}]{}

\section{Introduction}

Modeling the various transport phenomena in proton exchange membrane fuel cells (PEMFCs) with high accuracy can be a challenging task \cite{weber:14}, in particular at high current density. In spite of the well-known parametric degeneracy of fuel cell polarization curves, it is still common practice to validate PEMFC models with small sets of experimental performance data that cannot accommodate the full complexity of a fuel cell's nonlinear response to changes in operating conditions and material properties. It is all the more important that the uncertainties associated with each model parameter are known. This not only assists modelers with appreciating the limitations of their models, but also provides insight for fuel cell designers and engineers who seek to understand which material properties to tackle in order to increase cell performance or to reduce manufacturing costs.

Although a large number of numerical PEMFC models have been developed over the past decades, only few publications offer parameter sensitivity analyses \cite{siegel:08}. Kimble \& White \cite{kimble:92} presented an early parameter sensitivity study of a five-layer alkaline fuel cell model, focusing mainly on the effect of electrode thickness, porosity and interfacial surface area on the limiting current density. Reports of similar efforts for acidic fuel cell models seem to have appeared much more recently only. Guo et al.~\cite{guo:04} fitted five mass and charge transport parameters simultaneously to experimental polarization data, confirming the well-known significance of ionic conductivity and gas transport on PEMFC performance. Grujicic \& Chittajallu \cite{grujicic:04} reported on a flow channel design optimization with a coarsely meshed sensitivity analysis for six cathode parameters using a single-phase 2D model. Carnes \& Djilali \cite{carnes:05} estimated the local sensitivity of a simple 1D PEMFC model with three differential equations to changes in membrane conductivity, exchange current densities and oxygen diffusivities with a least-squares fit to experimental polarization curves. Min et al.~\cite{min:06} carried out a local sensitivity study with eleven varying parameters on a 3D two-phase PEMFC model and found that the cathode kinetic properties, the membrane conductivity and oxygen transport capability of the cathode GDL and are among the most significant ones. They also reemphasized the need to include more information than just the polarization curve in model validation. Mawardi \& Pitchumani \cite{mawardi:06} used latin hypercube sampling to examine the parameter uncertainty of a 1D single-phase PEMFC model. Multivariate uncertainty, in our parlance adopted here, refers to the presence of concurrent uncertainties in more than one input parameter. Their study was, however, limited to uncertainty in the kinetic transfer coefficients and the employed operating conditions rather than material parameterizations. Moreover, the variance of power density was the only criterion for uncertainty in the predicted fuel cell model response. Zhao et al.~\cite{zhao:15} performed a simple local variation of various parameters of a 1D PEMFC stack model. Laoun et al.~\cite{laoun:16} were, to the best of our knowledge, the first to report on a global multivariate sensitivity analysis of a PEMFC model. They drew quasi-random samples over a relatively broad range of operating conditions (gas pressure, temperature, current density) and some geometrical properties of the membrane-electrode assembly (MEA) sandwich (layer thicknesses, GDL porosity, cell area) to quantify the uncertainty of the predicted power density with Sobol indices. The research group found that their 0D model was most sensitive to variations in current density and membrane thickness. Several local parameter sensitivity analyses were also carried out on strongly simplified, lumped or zero-dimensional PEMFC models \cite{correa:05,placca:09,noorkami:14,noguer:15,correa:15} to examine the effect of uncertainty in one up to a few variables such as temperature or GDL porosity on fuel cell polarization. Of these, only the works by Corr\^{e}a et al. conclude with a list of model parameters sorted by sensitivity. Recently, Shah \cite{shah:17} proposed to use principal component analysis to reduce the computational demand associated with uncertainty quantification in PEMFC modeling with Monte Carlo sampling.

While the above-cited publications show a moderately increasing level of sophistication in PEMFC model uncertainty quantification, they are far from complete in the sense that:
\begin{itemize}
\item Only a very limited set of varying input parameters are considered, out of the dozens a high-fidelity PEM fuel cell model depends on.
\item Almost all studies examine only a single model output (typically the power density at a fixed operating point).
\item Parameter uncertainty is almost exclusively determined only \emph{locally}, i.e., at a fixed point of operation or model parameterization, rather than \emph{globally}, i.e., for a whole spectrum of conditions.
\item Most reported uncertainty analyses are based on computational models with strongly reduced complexity, and thus with insufficient account for the complex nonlinear multifunctional behavior of MEA materials.
\end{itemize}
Therefore, a thorough global parameter uncertainty quantification with a detailed PEMFC model with spatial resolution is appropriate. In the first part of this bipartite study \cite{vetter:p1}, we demonstrated how considerable epistemic uncertainty associated with measured material properties can induce a large spread in the predicted fuel cell performance. Being an extension of our one-dimensional two-phase MEA model \cite{vetter:18}, the model developed in Part I offers sufficient efficiency for a large number of numerical evaluations in manageable computation time without neglecting through-plane gradients, as lumped models do. In the present article, we employ our model to carry out the most extensive parameter uncertainty propagation analysis for PEM fuel cells reported in the scientific literature so far. Unlike existing studies, we include a dozen predicted fuel cell state variables in the uncertainty analysis (including performance and water balance characteristics), rather than considering a single point on the polarization curve as the only model response. In Sec.~\ref{sec:quantification}, a mathematical formalism is established, suitable for the quantification of local error propagation through a strongly nonlinear function such as the present MEA model. To establish some intuition for the model response to input uncertainty, a \emph{local} sensitivity analysis at typical fuel cell operating conditions is presented in Sec.~\ref{sec:local_sa}. The analysis is then extended to the \emph{global} scope by systematic quasi-stochastic variation of all relevant model parameters over a wide range of physically meaningful values in Sec.~\ref{sec:global_sa}. This allows us to deduce a robust sorted list of input parameters, ranking their absolute and relative significance for predictive state-of-the-art PEMFC modeling. We refer to \cite{vetter:p1} for all modeling details and adopt the nomenclature introduced therein.

\section{Quantification of uncertainty propagation}
\label{sec:quantification}

The most appropriate measure for the (local) propagation of uncertainty through a mathematical model or function is its condition number -- a concept that is routinely applied in linear and nonlinear algebra as well as numerical mathematics. Given a model function $\vec{f}\colon\mathbb{R}^n\to\mathbb{R}^m$, the local relative condition number of the $i$-th output value $f_i$ with respect to the $j$-th input variable $x_j$, at a given set of model parameters $\vec{x}$ with $f_i(\vec{x})\neq0$, is defined as \cite{trefethen:97}
\begin{equation}
\kappa_{ij}(\vec{x}) = \abs{\frac{\partial\log f_i}{\partial\log x_j}(\vec{x})} = \abs{\frac{x_j}{f_i(\vec{x})}\frac{\partial f_i}{\partial x_j}(\vec{x})}
\end{equation}
if $f_i$ is differentiable at $\vec{x}$. More generally, for a non-differentiable function,
\begin{equation}
\label{eq:rel_cond}
\kappa_{ij}(\vec{x}) = \lim_{\delta\to0\strut}\,\sup_{\abs{\Delta x_j}\leq\delta}\abs{\frac{x_j}{f_i(\vec{x})}\frac{f_i(\vec{x}+\Delta x_j\vec{e}_j)-f_i(\vec{x})}{\Delta x_j}},
\end{equation}
where $\vec{e}_j$ denotes the $j$-th standard unit vector and $\Delta x_j$ a small perturbation or error in the $j$-th model parameter.

The condition number measures the propagation of uncertainty through a model function $f$. A relative error in the $j$-th input variable is magnified by the factor $\kappa_{ij}$ in the output value $f_i$ (assuming no uncertainty in all others). A quantity of interest $f_i$ predicted by a model is sensitive to uncertainty in its parameter $x_j$ if $\kappa_{ij}$ is large, and conversely, insensitive if $\kappa_{ij}$ is small. When $f_i$ responds (locally) linearly to relative changes in $x_j$, one has $\kappa_{ij}=1$. In the following, by computing the relative condition number explicitly for a large range of operating conditions and material properties $\vec{x}$ and for numerous fuel cell state characteristics $f_i$, we show in which material parameterizations the largest source of uncertainty resides in macro-homogeneous two-phase fuel cell modeling.

In practice, Eq.~\ref{eq:rel_cond} needs to be approximated for numerical models. The most straightforward way to do this is by fixing $\delta$ to a very small constant value and setting $\Delta x_j=\delta x_j$, such that
\begin{equation}
\label{eq:rel_cond_approx}
\kappa_{ij}(\vec{x}) \approx \frac{1}{\delta}\abs{\frac{f_i(\vec{x}+\delta x_j\vec{e}_j)}{f_i(\vec{x})}-1}
\end{equation}
We use $\delta=10^{-3}$ in all subsequent calculations.

First, appropriate model output variables $f_i$ need to be selected. In order to cover a wide spectrum of potential properties of interest for typical PEMFC operation, we choose twelve key figures of which the former six are performance indicators derived from the predicted polarization curve, whereas the latter six characterize the state of heat and water balance within the fuel cell. The selected performance indicators are the limiting current density $I_\mathrm{max}$, the peak power density $P_\mathrm{max}$, the current densities $I_{0.8}$, $I_{0.6}$ and $I_{0.4}$ at applied cell voltages of 0.8\,V, 0.6\,V and 0.4\,V, respectively, and the cell voltage $U_1$ at 1\,A\,cm$^{-2}$. All these key properties are evaluated subject to the condition that they are attained under the simulated operating conditions and material parameterizations. Furthermore, at a fixed cell voltage of 0.6\,V, we select the peak temperature $T_\mathrm{max}$ across the MEA sandwich, the minimum local water content of the ionomer (including catalyst layers) $\lambda_\mathrm{min}$, the mean water content of the catalyst-coated membrane (CCM) \cite{vetter:18}
\begin{equation}
\lambda_\mathrm{avg}=\int_\mathrm{CCM}\epsilon_\mathrm{i}\lambda\diff x \bigg/ \int_\mathrm{CCM}\epsilon_\mathrm{i}\diff x,
\end{equation}
the ohmic resistance of the membrane
\begin{equation}
R_\mathrm{p} = \int_\mathrm{PEM}\frac{\diff x}{\sigma_\mathrm{p}},
\end{equation}
the maximum local liquid water saturation $s_\mathrm{max}$, and the net water flux across the membrane $j_\lambda$.

\begin{table*}[t]
	\centering
	\caption{Local model sensitivity to parameter uncertainty for the baseline model parameterization at the reference operating conditions listed in Tab.~\ref{tab:ref_opcond}. High sensitivities are highlighted with shaded cells, from white ($\log_{10}\kappa_{ij}\leq-1$) to dark gray ($\log_{10}\kappa_{ij}\geq0$).}
	\label{tab:sensitivity_local}
	\scalebox{\tabscale}{\begin{tabular}{l|rrrrrr|rrrrrr|r}
	\toprule
	& \multicolumn{13}{c}{$\log_{10}\kappa_{ij}$}\\
	\cmidrule{2-14}
	$x_j$\hspace{0.7em}\tikz\draw (0,0)--(7pt,-7pt);\hspace{0.7em}$f_i$ & \multicolumn{1}{c}{$I_\mathrm{max}$} & \multicolumn{1}{c}{$P_\mathrm{max}$} & \multicolumn{1}{c}{$I_{0.8}$} & \multicolumn{1}{c}{$I_{0.6}$} & \multicolumn{1}{c}{$I_{0.4}$} & \multicolumn{1}{c|}{$U_1$} & \multicolumn{1}{c}{$T_\mathrm{max}$} & \multicolumn{1}{c}{$\lambda_\mathrm{min}$} & \multicolumn{1}{c}{$\lambda_\mathrm{avg}$} & \multicolumn{1}{c}{$R_\mathrm{p}$} & \multicolumn{1}{c}{$s_\mathrm{max}$} & \multicolumn{1}{c|}{$j_\lambda$} & med.\\
	\midrule
	$P_\mathrm{A}=P_\mathrm{C}$ & \cc{-2.81} & \cc{-0.90} & \cc{0.03} & \cc{-0.71} & \cc{-1.36} & \cc{-0.99} & \cc{-2.00} & \cc{-0.52} & \cc{-0.74} & \cc{-0.35} & \cc{-1.44} & \cc{-0.82} & \cc{-0.86}\\
	$\mathrm{RH}_\mathrm{A}$ & \cc{-0.23} & \cc{-0.37} & \cc{-0.93} & \cc{-0.41} & \cc{-0.24} & \cc{-0.76} & \cc{-1.62} & \cc{0.01} & \cc{-0.21} & \cc{0.16} & \cc{-0.66} & \cc{0.56} & \cc{-0.30}\\
	$\mathrm{RH}_\mathrm{C}$ & \cc{-1.63} & \cc{-1.00} & \cc{-0.42} & \cc{-0.93} & \cc{-1.13} & \cc{-1.24} & \cc{-1.56} & \cc{-1.38} & \cc{-1.32} & \cc{-0.85} & \cc{-0.74} & \cc{-0.49} & \cc{-1.07}\\
	$T_\mathrm{A}=T_\mathrm{C}$ & \cc{0.07} & \cc{-0.24} & \cc{-0.09} & \cc{-0.43} & \cc{-0.05} & \cc{-0.82} & \cc{-0.01} & \cc{-0.02} & \cc{-0.29} & \cc{0.42} & \cc{-0.86} & \cc{0.16} & \cc{-0.07}\\
	$\alpha_{\mathrm{O}_2}$ & \cc{-1.47} & \cc{-0.84} & \cc{-0.31} & \cc{-0.77} & \cc{-1.05} & \cc{-1.07} & \cc{-2.20} & \cc{-0.72} & \cc{-0.94} & \cc{-0.54} & \cc{-1.83} & \cc{-1.38} & \cc{-0.99}\\
	\midrule
	$\sigma_\mathrm{e}$ & \cc{-2.93} & \cc{-1.71} & \cc{-1.57} & \cc{-1.58} & \cc{-2.05} & \cc{-1.88} & \cc{-2.99} & \cc{-1.52} & \cc{-1.74} & \cc{-1.33} & \cc{-2.69} & \cc{-2.23} & \cc{-1.81}\\
	$\sigma_\mathrm{p}$ & \cc{-1.95} & \cc{-0.70} & \cc{-0.95} & \cc{-0.69} & \cc{-0.90} & \cc{-1.02} & \cc{-2.17} & \cc{-0.62} & \cc{-0.80} & \cc{-0.21} & \cc{-1.76} & \cc{-1.01} & \cc{-0.92}\\
	$k$ & \cc{-1.59} & \cc{-1.96} & \cc{-2.70} & \cc{-2.07} & \cc{-1.67} & \cc{-2.42} & \cc{-1.87} & \cc{-1.40} & \cc{-1.62} & \cc{-1.42} & \cc{-1.69} & \cc{-0.90} & \cc{-1.68}\\
	$D_\lambda$ & \cc{-0.15} & \cc{-0.49} & \cc{-1.89} & \cc{-0.61} & \cc{-0.31} & \cc{-0.97} & \cc{-2.21} & \cc{-0.32} & \cc{-0.50} & \cc{-0.07} & \cc{-2.10} & \cc{-0.65} & \cc{-0.55}\\
	$\xi$ & \cc{0.25} & \cc{-0.06} & \cc{-1.43} & \cc{-0.17} & \cc{0.11} & \cc{-0.53} & \cc{-1.81} & \cc{0.12} & \cc{-0.05} & \cc{0.35} & \cc{-2.36} & \cc{-0.01} & \cc{-0.06}\\
	$M_\mathrm{p}$ & \cc{-1.33} & \cc{-0.92} & \cc{-1.31} & \cc{-0.93} & \cc{-1.01} & \cc{-1.26} & \cc{-2.71} & \cc{-1.30} & \cc{-1.46} & \cc{-1.04} & \cc{-2.05} & \cc{-0.64} & \cc{-1.28}\\
	$D_s$ & \cc{-2.09} & \cc{-3.97} & \cc{-1.84} & \cc{-2.98} & \cc{-2.46} & \cc{-3.19} & \cc{-3.03} & \cc{-1.90} & \cc{-1.86} & \cc{-1.59} & \cc{-0.90} & \cc{-2.55} & \cc{-2.28}\\
	$j_\mathrm{A}$ & \cc{-2.50} & \cc{-1.87} & \cc{-2.10} & \cc{-1.87} & \cc{-1.99} & \cc{-2.19} & \cc{-3.43} & \cc{-1.88} & \cc{-1.90} & \cc{-1.43} & --- & \cc{-1.70} & \cc{-1.94}\\
	$j_\mathrm{C}$ & \cc{-2.34} & \cc{-1.22} & \cc{-0.25} & \cc{-1.01} & \cc{-1.60} & \cc{-1.29} & \cc{-2.48} & \cc{-0.95} & \cc{-1.16} & \cc{-0.76} & \cc{-2.02} & \cc{-1.54} & \cc{-1.26}\\
	$\gamma_\mathrm{c}$ & \cc{-3.30} & \cc{-3.50} & \cc{-4.27} & \cc{-3.54} & \cc{-3.39} & \cc{-3.85} & \cc{-4.44} & \cc{-3.66} & \cc{-3.23} & \cc{-3.02} & \cc{-3.22} & \cc{-3.34} & \cc{-3.44}\\
	$\gamma_\mathrm{e}$ & \cc{-2.12} & \cc{-2.51} & \cc{-3.79} & \cc{-2.68} & \cc{-2.32} & \cc{-3.04} & \cc{-3.26} & \cc{-2.57} & \cc{-2.46} & \cc{-2.25} & \cc{-2.03} & \cc{-2.35} & \cc{-2.48}\\
	$k_\mathrm{a}$ & \cc{-0.79} & \cc{-1.04} & \cc{-2.99} & \cc{-1.16} & \cc{-0.86} & \cc{-1.52} & \cc{-2.44} & \cc{-0.84} & \cc{-1.00} & \cc{-0.64} & \cc{-1.53} & \cc{-0.31} & \cc{-1.02}\\
	$k_\mathrm{d}$ & \cc{-1.66} & \cc{-1.76} & \cc{-2.76} & \cc{-1.81} & \cc{-1.70} & \cc{-2.15} & \cc{-3.38} & \cc{-1.87} & \cc{-1.49} & \cc{-1.32} & \cc{-2.84} & \cc{-1.91} & \cc{-1.84}\\
	$R_\mathrm{e}$ & \cc{-2.33} & \cc{-0.80} & \cc{-0.65} & \cc{-0.67} & \cc{-1.16} & \cc{-0.97} & \cc{-2.00} & \cc{-0.60} & \cc{-0.81} & \cc{-0.42} & \cc{-1.85} & \cc{-1.44} & \cc{-0.89}\\
	$R_T$ & \cc{-1.25} & \cc{-1.67} & \cc{-2.67} & \cc{-1.78} & \cc{-1.43} & \cc{-2.14} & \cc{-1.79} & \cc{-1.26} & \cc{-1.35} & \cc{-1.16} & \cc{-1.39} & \cc{-0.73} & \cc{-1.41}\\
	$\lambda_\mathrm{v}$ & \cc{-0.24} & \cc{-0.37} & \cc{-0.90} & \cc{-0.38} & \cc{-0.29} & \cc{-0.72} & \cc{-1.78} & \cc{-0.20} & \cc{-0.13} & \cc{0.14} & \cc{-1.03} & \cc{0.12} & \cc{-0.33}\\
	$\epsilon_\mathrm{i}^\mathrm{CL}$ & \cc{-0.69} & \cc{-0.78} & \cc{-1.29} & \cc{-0.82} & \cc{-0.75} & \cc{-1.16} & \cc{-2.33} & \cc{-1.10} & \cc{-0.87} & \cc{-0.57} & \cc{-1.89} & \cc{-1.29} & \cc{-0.98}\\
	$L_0^\mathrm{GDL}$ & \cc{-1.16} & \cc{-0.83} & \cc{-1.07} & \cc{-0.82} & \cc{-0.90} & \cc{-1.14} & \cc{-1.97} & \cc{-1.35} & \cc{-1.47} & \cc{-0.95} & \cc{-1.05} & \cc{-0.41} & \cc{-1.06}\\
	$L_0^\mathrm{CL}$ & \cc{-0.73} & \cc{-1.04} & \cc{-0.29} & \cc{-1.39} & \cc{-0.85} & \cc{-1.90} & \cc{-3.17} & \cc{-0.70} & \cc{-0.77} & \cc{-0.33} & \cc{-2.19} & \cc{-1.34} & \cc{-0.95}\\
	$L_0^\mathrm{PEM}$ & \cc{-0.29} & \cc{-0.45} & \cc{-1.13} & \cc{-0.53} & \cc{-0.37} & \cc{-0.88} & \cc{-2.16} & \cc{-0.82} & \cc{-1.20} & \cc{0.07} & \cc{-1.87} & \cc{-0.73} & \cc{-0.77}\\
	$P_\mathrm{cl}$ & \cc{-1.06} & \cc{-0.81} & \cc{-1.16} & \cc{-0.72} & \cc{-0.97} & \cc{-1.03} & \cc{-2.20} & \cc{-0.95} & \cc{-1.32} & \cc{-0.83} & \cc{-1.05} & \cc{-0.76} & \cc{-1.00}\\
	\midrule
	median & \cc{-1.40} & \cc{-0.91} & \cc{-1.23} & \cc{-0.87} & \cc{-1.03} & \cc{-1.20} & \cc{-2.20} & \cc{-0.95} & \cc{-1.18} & \cc{-0.70} & \cc{-1.84} & \cc{-0.86} & \cc{-1.01}\\
	\bottomrule
	\end{tabular}}
\end{table*}

Next, a set of model input factors $x_j$ are chosen, with respect to which the modeling uncertainty is to be evaluated. They can be classified into two primary categories: the conditions at which the fuel cell is operated, and the MEA properties including electrochemical, physical and geometrical parameters. In the first category we have the anode and cathode gas feed pressures $P_\mathrm{A}$ and $P_\mathrm{C}$, the relative humidities $\mathrm{RH}_\mathrm{A}$ and $\mathrm{RH}_\mathrm{C}$ in the gas channels, the boundary temperatures $T_\mathrm{A}$ and $T_\mathrm{C}$ (assuming that the temperature of the supplied gas equals that of the bipolar plate on either side of the MEA) and the mole fraction of oxygen in the oxidation gas mixture, which equals 21\% for operation with ambient air. The parameter set in the second category consists of the following coefficients: The electric conductivity of the catalyst layers (CLs) and gas diffusion layers (GDLs) $\sigma_\mathrm{e}$, the protonic conductivity of the ionomer $\sigma_\mathrm{p}$, the thermal conductivity $k$, the effective Fickean diffusivity of water in the ionomer $D_\lambda$, the electro-osmotic drag (EOD) coefficient $\xi$, the microstructure factor of the pores in the CLs and GDLs $M_\mathrm{p}$ that reduces the effective gas diffusivities, the liquid water transport coefficient $D_s$ (incorporating the absolute and relative hydraulic permeabilities but also the capillary pressure--saturation relationship), the anode and cathode exchange current densities $j_\mathrm{A}$ and $j_\mathrm{C}$, the condensation and evaporation rates $\gamma_\mathrm{c}$ and $\gamma_\mathrm{e}$, the vapor absorption and desorption rates at the ionomer--gas interface $k_\mathrm{a}$ and $k_\mathrm{d}$, the electrical and thermal contact resistivities $R_\mathrm{e}$ and $R_T$, the hydration number of a vapor-equilibrated membrane $\lambda_\mathrm{v}$, the volume fraction of ionomer in the catalyst layer $\epsilon_\mathrm{i}^\mathrm{CL}$, the uncompressed layer thicknesses $L_0^\mathrm{GDL}$, $L_0^\mathrm{CL}$ and $L_0^\mathrm{PEM}$, and finally, the applied clamping pressure $P_\mathrm{cl}$, which affects several effective MEA parameters simultaneously.

\section{Local sensitivity analysis}
\label{sec:local_sa}

\begin{table}
	\centering
	\caption{Reference operating conditions (from \cite{vetter:p1}).\strut}
	\label{tab:ref_opcond}
	\scalebox{\tabscale}{\begin{tabular}{llr}
	\toprule
	Symbol & Explanation & Value\\
	\midrule
	$P_\mathrm{A}$ & Gas pressure in anode gas channel & $1.5\,\mathrm{bar}$\\
	$P_\mathrm{C}$ & Gas pressure in cathode gas channel & $1.5\,\mathrm{bar}$\\
	$\mathrm{RH}_\mathrm{A}$ & Relative humidity in anode gas channel & $100\%$\\
	$\mathrm{RH}_\mathrm{C}$ & Relative humidity in cathode gas channel & $100\%$\\
	$T_\mathrm{A}$ & Temperature of anode plate and gas channel & $80\,^\circ\mathrm{C}$\\
	$T_\mathrm{C}$ & Temperature of cathode plate and gas channel & $80\,^\circ\mathrm{C}$\\
	$\alpha_{\mathrm{O}_2}$ & Oxygen mole fraction in dry oxidant gas & $21\%$\\
	\bottomrule
	\end{tabular}}
\end{table}

\begin{figure*}[t]
	\centering
	\includegraphics{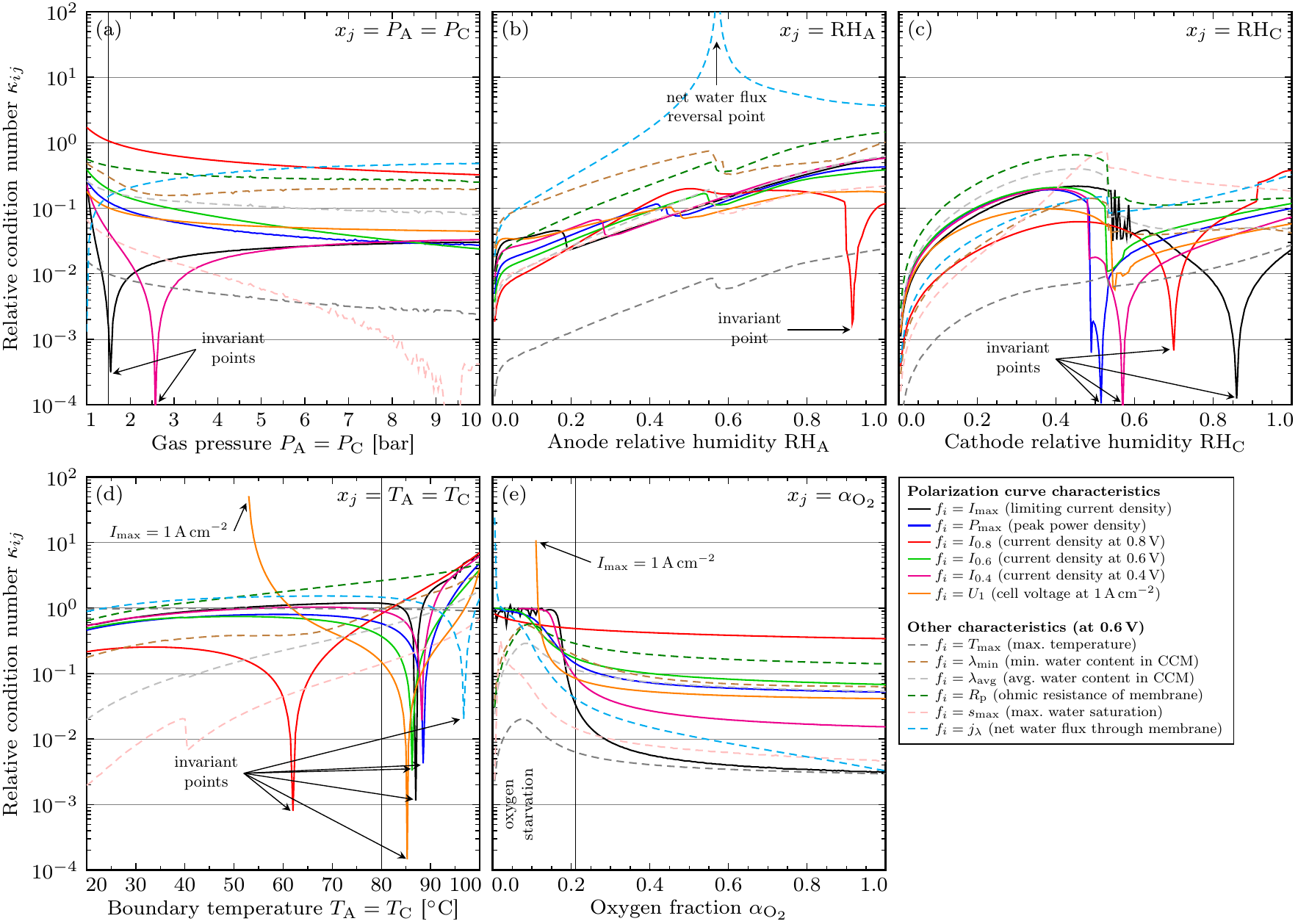}
	\caption{Model sensitivity to uncertainty in the operating conditions for the baseline model parameterization. All operating conditions are fixed according to Tab.~\ref{tab:ref_opcond} but the one shown on the respective horizontal axes. Thin solid vertical lines indicate the reference conditions. Local fluctuations are numerical noise.}
	\label{fig:local_cond_opcond}
\end{figure*}

\begin{figure*}
	\centering
	\includegraphics{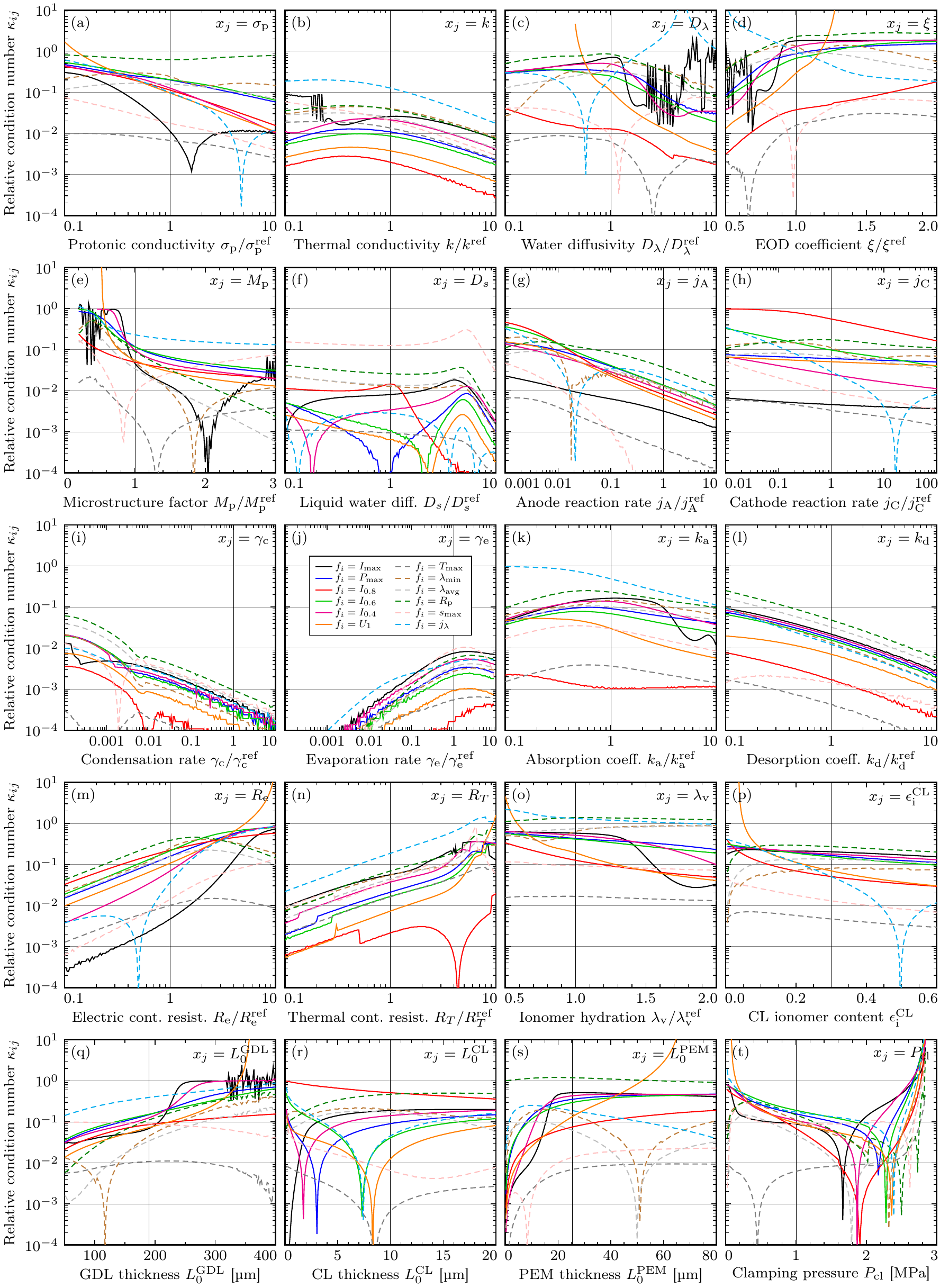}
	\caption{Model sensitivity to uncertainty in the parameterization at the reference operating conditions as listed in Tab.~\ref{tab:ref_opcond}. All parameterizations are fixed but the one shown on the respective horizontal axes. The graphs represent the same output variables $f_i$ as in Fig.~\ref{fig:local_cond_opcond}. Thin solid vertical lines indicate the baseline parameterization. Local fluctuations are numerical noise.}
	\label{fig:local_cond_param}
\end{figure*}

\begin{figure}
	\centering
	\includegraphics{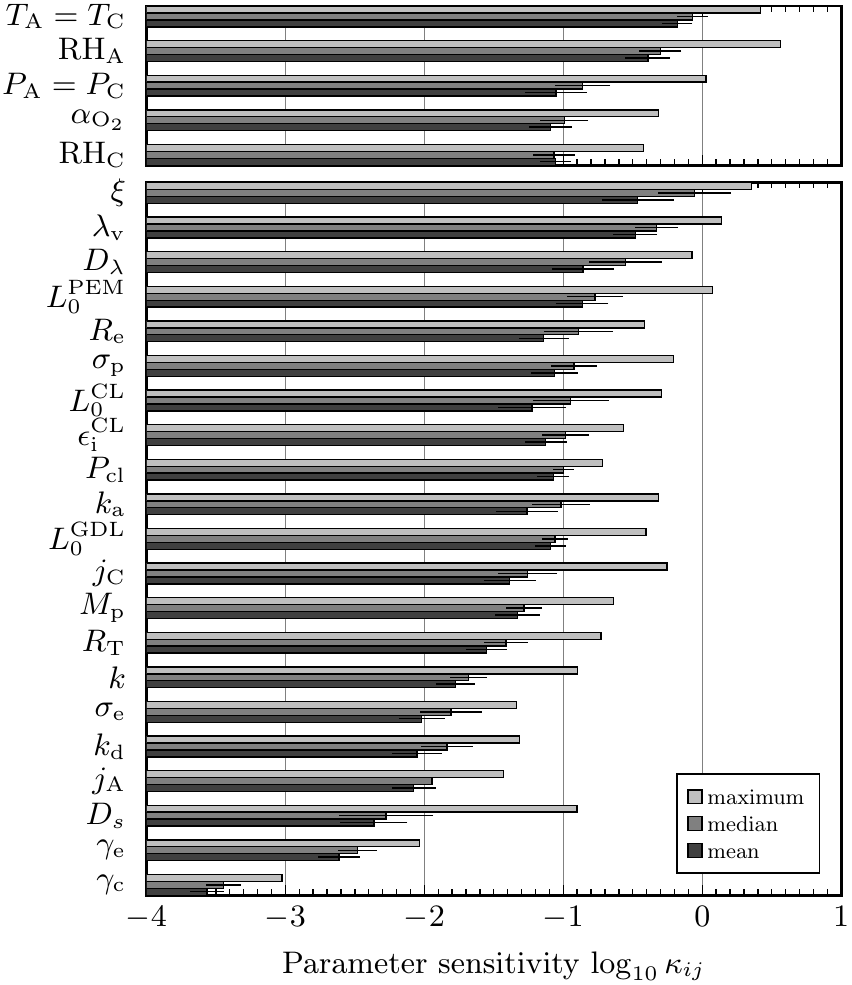}
	\caption{Local model sensitivity to parameter uncertainty for the baseline model parameterization at the reference operating conditions. Operating conditions (top group) and MEA parameterizations (bottom group) are separately sorted by significance (i.e., by median of $\log_{10}\kappa_{ij}$ taken over all outputs $i$) in decreasing order. Error bars are standard errors of the statistics over $i$.}
	\label{fig:sensitivity_local}
\end{figure}

We start off with a local sensitivity analysis to showcase the model behavior using the best material parameterization found in Part I \cite{vetter:p1}, at operating conditions typical for automotive applications.

Many of the material parameters are based on different constitutive parameterizations in the individual MEA layers, as detailed in Part I. The aim of the present study is to compare the impact of the different kinds of model parameters rather than the impact of individual layers. We therefore vary each parameter $x_j$ in all layers simultaneously where applicable. As an example, if the $j$-th model parameter is the thermal conductivity $k$, we perturb $k$ in \emph{each} MEA layer by the factor $\delta$ for the evaluation of $\kappa_{ij}$ using Eq.~\ref{eq:rel_cond_approx}. Furthermore, we set $P_\mathrm{A}=P_\mathrm{C}$ and $T_\mathrm{A}=T_\mathrm{C}$ to reduce the number of independent operating conditions from seven to five, which simplifies the data presentation.

Since the condition number is a logarithmically scaling quantity, we report all values in logarithmic form. A fuel cell characteristic $f_i$ is highly sensitive to uncertainty in parameter $x_j$ if $\log_{10}\kappa_{ij}$ is positive or nearly so. Tab.~\ref{tab:sensitivity_local} lists the simulation results for all input/output pairs at the reference operating conditions given in Tab.~\ref{tab:ref_opcond}. Note that the maximum liquid water saturation on the cathode side is insensitive to changes in the anode reaction rate (i.e., the condition number is indistinguishable from zero within the numerical error tolerances used here), which is why this value is omitted in Tab.~\ref{tab:sensitivity_local}. This data provides first insight into the model behavior at reference operating conditions. The peak temperature and the maximum liquid water saturation are the least sensitive of the twelve selected traits quantifying the operative state of the fuel cell. Nevertheless, with the membrane resistance and the net water flux, two quantities from the second category (not derived from the polarization curve) are the most sensitive overall, which accentuates that a comprehensive uncertainty analysis should also include model outputs other than points on the polarization curve.

The relative humidity in the anode gas channel and the boundary temperature are the most influential operating conditions at reference conditions, which can be explained by the direct impact on the ionic conductivity on the (typically drier) anode side of the membrane they have \cite{vetter:18,vetter:p1}. In the category of MEA parameters, the water diffusivity, the EOD coefficient and the equilibrium water uptake of the ionomer show the greatest impact on many of the fuel cell properties predicted by the model.

How sensitive these findings are with respect to the choice of operating conditions is demonstrated in Fig.~\ref{fig:local_cond_opcond}, where in each subplot one of them is systematically varied. All model parameters are fixed but the one operating factor with respect to which the condition number is evaluated. As can be recognized from Fig.~\ref{fig:local_cond_opcond}a, changes in gas pressure mostly affect the polarization at low current densities ($I_{0.8}$), whereas the maximum temperature and liquid water saturations are generally very insensitive. An important feature of the relative condition number is that its logarithm is unbounded toward both negative and positive infinity. $\kappa_{ij}$ vanishes if $x_j=0$ or $\partial f_i/\partial x_j=0$. Since we varied the gas pressure from 1 to 10\,bar, only the second condition can be fulfilled in Fig.~\ref{fig:local_cond_opcond}a. Indeed, the two pressure points at which $\kappa_{ij}\to0$, the current densities $I_{0.4}$ and $I_\mathrm{max}$ are invariants on the polarization curve. For instance, if the gas pressure is varied about the value of 2.6\,bar, the resulting polarization curves cross at 0.4\,V, because that is where $\log_{10}\kappa_{ij}=-\infty$.

The reverse scenario occurs with the net water flux through the membrane $j_\lambda$ in Fig.~\ref{fig:local_cond_opcond}b. A high relative humidity is required in the hydrogen gas feed to obtain a flux from anode to cathode. Slightly below $\mathrm{RH}_\mathrm{A}=60\%$ (for fixed $\mathrm{RH}_\mathrm{C}=100\%$), back diffusion and EOD cancel each other, such that $f_i=j_\lambda=0$, letting $\kappa_{ij}$ diverge to infinity. At lower $\mathrm{RH}_\mathrm{A}$, the hydration gradient in the membrane is so steep that back diffusion is stronger than EOD at 0.6\,V. This reversal of water balance also affects other fuel cell properties, which show a kink at that relative humidity value, such as $T_\mathrm{max}$ (through the latent heat of phase change and sorption) or the ohmic membrane resistance $R_\mathrm{p}$ (through the strong hydration-dependence of the ionic conductivity).

There is also an invariant point on the polarization curve at 0.8\,V slightly above $\mathrm{RH}_\mathrm{A}=90\%$ (Fig.~\ref{fig:local_cond_opcond}b), in the vicinity of which $\partial I_{0.8}/\partial\mathrm{RH}_\mathrm{A}\approx0$. As $\mathrm{RH}_\mathrm{A}\to0$, also the relative condition numbers vanish. The model sensitivity to uncertainty in the RH on the anode side generally increases toward higher humidity. A similar observation is made in Fig.~\ref{fig:local_cond_opcond}c for $\mathrm{RH}_\mathrm{C}$ until it reaches about 50\%, above which the model abruptly looses its sensitivity to RH changes in the cathode gas channel. This is because with a more humid air feed, evaporation yields saturated gas in almost the entire CL and GDL, making further humidification of the supplied air irrelevant. Only as $\mathrm{RH}_\mathrm{C}\to100\%$, the sensitivity rises again to significant values, because liquid water starts to accumulate.

As shown in Fig.~\ref{fig:local_cond_opcond}d, various points on the polarization curve are invariant points with respect to changes in the boundary temperature at different temperature values, which can be rationalized by appreciating how many material parameterizations in our model are temperature-dependent. Also the net water flux is found to be invariant at about 97\,$^\circ$C. As expected, the effect of the boundary temperature on $T_\mathrm{max}$ is nearly linear, as can be recognized by $\kappa_{ij}\approx1$ in Fig.~\ref{fig:local_cond_opcond}d. Low temperatures prevent the fuel cell from operating at higher current densities, which is why the condition number for $U_1$ diverges when the temperature is such that $I_\mathrm{max}=1\,\mathrm{A\,cm}^{-2}$, i.e., at about 53\,$^\circ$C. The same effect is observed in Fig.~\ref{fig:local_cond_opcond}e for the case of oxygen starvation ($\alpha_{\mathrm{O}_2}\lesssim11\%$). Not surprisingly, polarization curve features at large current densities ($I_\mathrm{max}$, $I_{0.4}$, $I_{0.6}$, $P_\mathrm{max}$) exhibit $\kappa_{ij}\approx1$ as oxygen depletes, which means that $I$ responds linearly to changes in $\alpha_{\mathrm{O}_2}$ in this oxygen-limited regime.

Next, the same analysis is repeated for the 21 major MEA parameters to obtain a quantitative picture of how the model sensitivity responds to deviations from the baseline parameterization. For those which are parametric expressions or layer-dependent properties rather than constant coefficients, we introduce a prefactor $C$ to be varied, while maintaining the functional dependency of the constitutive parameterization. Consider, for instance, the ionic conductivity of the ionomer $\sigma_\mathrm{p}$. We set $\sigma_\mathrm{p}=C\sigma_\mathrm{p}^\mathrm{ref}$, where
\begin{equation}
\sigma_\mathrm{p}^\mathrm{ref} = M_\mathrm{i} \sigma_0 \max\{f_\mathrm{w}-f_0\}^\beta
\end{equation}
is the baseline parameterization from Part I \cite{vetter:p1}. Thus, the prefactor $C=A/A^\mathrm{ref}$ is reported in Fig.~\ref{fig:local_cond_param} for $A=\sigma_\mathrm{p},k,D_\lambda,\xi,M_\mathrm{p},D_s,j_\mathrm{A},j_\mathrm{C},\gamma_\mathrm{c},\gamma_\mathrm{e},k_\mathrm{a},k_\mathrm{d},R_\mathrm{e},R_T,\lambda_\mathrm{v}$. We omit the data for the electric conductivity of the CLs and GDLs ($\sigma_\mathrm{e}$), which is one of the least interesting, to save space. In short, we find the expected behavior $\kappa_{ij}\sim(\sigma_\mathrm{e}/\sigma_\mathrm{e}^\mathrm{ref})^{-1}$ for all $i$ in the examined range $0.1\leq\sigma_\mathrm{e}/\sigma_\mathrm{e}^\mathrm{ref}\leq10$ with prefactors as listed in Tab.~\ref{tab:sensitivity_local}, which is generally very low. We restrict the discussion of Fig.~\ref{fig:local_cond_param} to a few relevant points:
\begin{itemize}
\item Notice that relatively broad parameter ranges are screened -- broader than the scatter of available experimental data for some of them \cite{vetter:p1}. A few of the more extreme transport coefficients yield noisy sensitivity data, e.g., large values for $D_\lambda$ (Fig.~\ref{fig:local_cond_param}c).
\item Our baseline parameterizations for the anode exchange current density and the evaporation/condensation rates are near the upper end of the reported experimental range, which is why their domains analyzed here extend further down than up.
\item Invariant points (zeros of $\kappa_{ij}$) are scattered all over the parameter domains.
\item A water flux reversal point is present at high back diffusivities. At low back diffusivity, large EOD coefficients, low gas diffusivity, high electrical contact resistivity, poor membrane hydration, small ionomer content in the CLs, thick GDLs and membranes, and at high clamping pressures, the limiting current density drops below $1\,\mathrm{A\,cm}^{-2}$, resulting in poles for $U_1$.
\item $T_\mathrm{max}$ and $s_\mathrm{max}$ are often the most insensitive features, also away from the baseline parameterization. However, it is not possible to identify a single quantity $f_i$ which could serve as a robust representative for the overall model sensitivity.
\end{itemize}

Evidently, it is essential to look at more than a single point in the parameter space and also at more than one output variable for a robust and generally valid evaluation of parameter uncertainty in a PEMFC model, because one might accidentally hit a spot where $\kappa_{ij}$ is near zero or extremely large. A \emph{global} parameter sensitivity analysis is therefore indispensable. Moreover, since zeros and poles will be encountered even in a global analysis, an appropriate robust statistic should be chosen to quantitatively compare the impact of different model parameters on the model prediction. The extrema (minimum and maximum) over all $i$ or $j$ will be of very limited information value, as they strongly depend on the choice of $f_i$ and the screened range of admissible input values $x_j$. We therefore choose the median of $\log_{10}\kappa_{ij}$ (over $i$ or $j$), which has a breakdown point of 50\%, as the decisive statistic to quantify uncertainty. The calculated medians of the local sensitivity analysis are given in Tab.~\ref{tab:sensitivity_local}.

Fig.~\ref{fig:sensitivity_local} summarizes the results of the local sensitivity analysis. The two input parameter categories are sorted by median over all outputs $i$. Alongside with the medians, the maximum and mean values over $i$ are also given for comparative purposes. At reference automotive conditions with the baseline parameterization, the boundary temperature and the anode feed humidity are the most influential operating conditions. In the category of MEA properties, the ability of the ionomer to accommodate, transport and spatially redistribute dissolved water clearly dominates, followed by the membrane thickness and electrical contact resistivity. This corroborates our previous recommendation that contact resistance should not be neglected in PEMFC modeling \cite{vetter:17}.

\section{Global sensitivity analysis}
\label{sec:global_sa}

\begin{table*}[p]
	\centering
	\caption{Parameter ranges for the global sensitivity analysis. For MEA parameterizations with functional dependencies on other variables, the functional dependency is fixed and only the prefactor $C$ is varied (denoted by $C=A/A^\mathrm{ref}$, where $A$ is the effective resulting value of the parameter and $A^\mathrm{ref}$ the baseline parameterization as detailed in Part I). In these cases, the indicated range of values applies to this prefactor.}
	\label{tab:test_intervals}
	\scalebox{\tabscale}{\begin{tabular}{cllllll}
	\toprule
	& Parameter & Explanation & Unit & Range of values & Scaling in range & Details in Part I\\
	\midrule[\heavyrulewidth]
	\parbox[t]{2em}{\multirow{5}{*}{\rotatebox[origin=c]{90}{\parbox{5\baselineskip}{\centering operating\\conditions}}}} & $P_\mathrm{A}=P_\mathrm{C}$ & Gas feed pressure & bar & $1-4$ & linear & Sec.~2.4\\
	& $\mathrm{RH}_\mathrm{A}$ & Relative humidity in anode gas channel & --- & $0.5-1$ & linear & Sec.~2.4\\
	& $\mathrm{RH}_\mathrm{C}$ & Relative humidity in cathode gas channel & --- & $0.5-1$ & linear & Sec.~2.4\\
	& $T_\mathrm{A}=T_\mathrm{C}$ & Temperature of bipolar plates & $^\circ$C & $50-90$ & linear & Sec.~2.4\\
	& $\alpha_{\mathrm{O}_2}$ & Oxygen mole fraction in dry oxidant gas & --- & $0.1-0.3$ & linear & Sec.~2.4\\
	\midrule
	\parbox[t]{2em}{\multirow{21}{*}{\rotatebox[origin=c]{90}{MEA parameterizations}}} & $\sigma_\mathrm{e}/\sigma_\mathrm{e}^\mathrm{ref}$ & Electric conductivity of CLs and GDLs & --- & $0.1-10$ & logarithmic & Sec.~3.2\\
	& $\sigma_\mathrm{p}/\sigma_\mathrm{p}^\mathrm{ref}$ & Protonic conductivity of the ionomer & --- & $0.1-10$ & logarithmic & Eq.~(22)\\
	& $k/k^\mathrm{ref}$ & Thermal conductivity & --- & $0.1-10$ & logarithmic & Eqs.~(25)--(32)\\
	& $D_\lambda/D_\lambda^\mathrm{ref}$ & Diffusivity of water in the ionomer & --- & $0.1-10$ & logarithmic & Eqs.~(34), (35), last row of Tab.~3\\
	& $\xi/\xi^\mathrm{ref}$ & Electro-osmotic drag coefficient & --- & $0.5-2$ & linear & Eqs.~(36), (38)\\
	& $M_\mathrm{p}/M_\mathrm{p}^\mathrm{ref}$ & Microstructure factor of the CLs and GDLs & --- & $0.2-3$ & linear & Eq.~(48)\\
	& $D_s/D_s^\mathrm{ref}$ & Liquid water transport coefficient & --- & $0.1-10$ & logarithmic & Eq.~(15)\\
	& $j_\mathrm{A}/j_\mathrm{A}^\mathrm{ref}$ & Anode exchange current density & --- & $0.001-10$ & logarithmic & Eq.~(21)\\
	& $j_\mathrm{C}/j_\mathrm{C}^\mathrm{ref}$ & Cathode exchange current density & --- & $0.01-100$ & logarithmic & Eq.~(21)\\
	& $\gamma_\mathrm{c}/\gamma_\mathrm{c}^\mathrm{ref}$ & Condensation rate & --- & $0.0001-10$ & logarithmic & last row of Tab.~7\\
	& $\gamma_\mathrm{e}/\gamma_\mathrm{e}^\mathrm{ref}$ & Evaporation rate & --- & $0.0001-10$ & logarithmic & last row of Tab.~7\\
	& $k_\mathrm{a}/k_\mathrm{a}^\mathrm{ref}$ & Vapor absorption rate of the ionomer & --- & $0.1-10$ & logarithmic & 4\textsuperscript{th} row of Tab.~6\\
	& $k_\mathrm{d}/k_\mathrm{d}^\mathrm{ref}$ & Vapor desorption rate of the ionomer & --- & $0.1-10$ & logarithmic & 4\textsuperscript{th} row of Tab.~6\\
	& $R_\mathrm{e}/R_\mathrm{e}^\mathrm{ref}$ & Electrical contact resistivity & --- & $0.1-10$ & logarithmic & Eq.~(68), Tab.~8\\
	& $R_T/R_T^\mathrm{ref}$ & Thermal contact resistivity & --- & $0.1-10$ & logarithmic & Eq.~(68), Tab.~8\\
	& $\lambda_\mathrm{v}/\lambda_\mathrm{v}^\mathrm{ref}$ & Vapor sorption isotherm of the ionomer & --- & $0.5-2$ & linear & Eqs.~(44)--(46)\\
	& $\epsilon_\mathrm{i}^\mathrm{CL}$ & Volume fraction of ionomer in the CLs & --- & $0.1-0.5$ & linear & Eq.~(23)\\
	& $L_0^\mathrm{GDL}$ & Uncompressed GDL thickness & \textmu{}m & $100-400$ & linear & Eqs.~(61), (62)\\
	& $L_0^\mathrm{CL}$ & Uncompressed CL thickness & \textmu{}m & $1-20$ & linear & Eqs.~(61), (63)\\
	& $L_0^\mathrm{PEM}$ & Uncompressed membrane thickness & \textmu{}m & $10-80$ & linear & Sec.~3.13\\
	& $P_\mathrm{cl}$ & Applied clamping pressure & MPa & $0.1-3$ & linear & Sec.~3.13\\
	\bottomrule
	\end{tabular}}
\end{table*}

\begin{table*}[p]
	\centering
	\caption{Global model sensitivity to parameter uncertainty. High sensitivities are highlighted with shaded cells, from white ($\log_{10}\kappa_{ij}\leq-1$) to dark gray ($\log_{10}\kappa_{ij}\geq0$). Standard errors (calculated with the bootstrap method, not shown) are in the second to fourth significant digit.}
	\label{tab:sensitivity_global}
	\scalebox{\tabscale}{\begin{tabular}{l|rrrrrr|rrrrrr|r}
	\toprule
	& \multicolumn{13}{c}{$\log_{10}\kappa_{ij}$}\\
	\cmidrule{2-14}
	$x_j$\hspace{0.7em}\tikz\draw (0,0)--(7pt,-7pt);\hspace{0.7em}$f_i$ & \multicolumn{1}{c}{$I_\mathrm{max}$} & \multicolumn{1}{c}{$P_\mathrm{max}$} & \multicolumn{1}{c}{$I_{0.8}$} & \multicolumn{1}{c}{$I_{0.6}$} & \multicolumn{1}{c}{$I_{0.4}$} & \multicolumn{1}{c|}{$U_1$} & \multicolumn{1}{c}{$T_\mathrm{max}$} & \multicolumn{1}{c}{$\lambda_\mathrm{min}$} & \multicolumn{1}{c}{$\lambda_\mathrm{avg}$} & \multicolumn{1}{c}{$R_\mathrm{p}$} & \multicolumn{1}{c}{$s_\mathrm{max}$} & \multicolumn{1}{c|}{$j_\lambda$} & med.\\
	\midrule
$P_\mathrm{A}=P_\mathrm{C}$ & \cc{-0.19} & \cc{-0.33} & \cc{-0.58} & \cc{-0.29} & \cc{-0.28} & \cc{-0.85} & \cc{-1.97} & \cc{-0.12} & \cc{-0.34} & \cc{0.14} & \cc{-1.06} & \cc{0.50} & \cc{-0.31}\\
$\mathrm{RH}_\mathrm{A}$ & \cc{-0.91} & \cc{-1.07} & \cc{-1.00} & \cc{-1.01} & \cc{-1.06} & \cc{-1.41} & \cc{-2.24} & \cc{-1.22} & \cc{-0.47} & \cc{-0.32} & \cc{-1.28} & \cc{-0.67} & \cc{-1.03}\\
$\mathrm{RH}_\mathrm{C}$ & \cc{-0.98} & \cc{-0.94} & \cc{-0.27} & \cc{-0.78} & \cc{-0.98} & \cc{-0.91} & \cc{-2.46} & \cc{-1.06} & \cc{-1.12} & \cc{-0.67} & \cc{-1.56} & \cc{-0.61} & \cc{-0.96}\\
$T_\mathrm{A}=T_\mathrm{C}$ & \cc{0.02} & \cc{-0.16} & \cc{-0.40} & \cc{-0.22} & \cc{-0.08} & \cc{-0.78} & \cc{-0.00} & \cc{-0.75} & \cc{-0.69} & \cc{0.11} & \cc{-1.16} & \cc{0.21} & \cc{-0.19}\\
$\alpha_{\mathrm{O}_2}$ & \cc{-1.86} & \cc{-1.20} & \cc{-0.49} & \cc{-1.03} & \cc{-1.45} & \cc{-1.04} & \cc{-2.82} & \cc{-1.47} & \cc{-1.69} & \cc{-1.02} & \cc{-2.38} & \cc{-1.33} & \cc{-1.39}\\
\midrule
$\sigma_\mathrm{e}$ & \cc{-2.87} & \cc{-2.10} & \cc{-2.13} & \cc{-2.23} & \cc{-2.61} & \cc{-1.64} & \cc{-3.84} & \cc{-2.51} & \cc{-2.71} & \cc{-2.09} & \cc{-3.46} & \cc{-2.40} & \cc{-2.45}\\
$\sigma_\mathrm{p}$ & \cc{-1.01} & \cc{-0.75} & \cc{-0.86} & \cc{-0.80} & \cc{-0.96} & \cc{-0.85} & \cc{-2.57} & \cc{-0.96} & \cc{-1.20} & \cc{-0.10} & \cc{-2.05} & \cc{-0.91} & \cc{-0.93}\\
$k$ & \cc{-1.75} & \cc{-2.14} & \cc{-3.15} & \cc{-2.26} & \cc{-2.05} & \cc{-2.11} & \cc{-2.34} & \cc{-2.01} & \cc{-2.09} & \cc{-1.81} & \cc{-2.61} & \cc{-1.59} & \cc{-2.10}\\
$D_\lambda$ & \cc{-0.27} & \cc{-0.56} & \cc{-1.81} & \cc{-0.80} & \cc{-0.50} & \cc{-1.32} & \cc{-2.65} & \cc{-0.91} & \cc{-0.87} & \cc{-0.30} & \cc{-1.91} & \cc{-0.50} & \cc{-0.83}\\
$\xi$ & \cc{0.21} & \cc{-0.06} & \cc{-1.62} & \cc{-0.25} & \cc{0.08} & \cc{-1.02} & \cc{-2.19} & \cc{-0.35} & \cc{-0.53} & \cc{0.23} & \cc{-1.50} & \cc{-0.09} & \cc{-0.30}\\
$M_\mathrm{p}$ & \cc{-0.94} & \cc{-1.07} & \cc{-1.71} & \cc{-1.19} & \cc{-1.09} & \cc{-1.33} & \cc{-2.96} & \cc{-1.29} & \cc{-1.19} & \cc{-0.88} & \cc{-1.93} & \cc{-0.56} & \cc{-1.19}\\
$D_s$ & \cc{-2.16} & \cc{-2.37} & \cc{-3.20} & \cc{-2.49} & \cc{-2.38} & \cc{-2.36} & \cc{-3.89} & \cc{-2.60} & \cc{-2.09} & \cc{-1.91} & \cc{-0.93} & \cc{-2.31} & \cc{-2.36}\\
$j_\mathrm{A}$ & \cc{-2.04} & \cc{-1.57} & \cc{-1.73} & \cc{-1.73} & \cc{-1.94} & \cc{-1.38} & \cc{-3.38} & \cc{-1.97} & \cc{-2.04} & \cc{-1.30} & \cc{-2.88} & \cc{-1.58} & \cc{-1.83}\\
$j_\mathrm{C}$ & \cc{-2.09} & \cc{-1.25} & \cc{-0.46} & \cc{-1.12} & \cc{-1.70} & \cc{-1.20} & \cc{-2.93} & \cc{-1.50} & \cc{-1.71} & \cc{-1.02} & \cc{-2.43} & \cc{-1.37} & \cc{-1.44}\\
$\gamma_\mathrm{c}$ & \cc{-3.64} & \cc{-3.98} & --- & \cc{-3.83} & \cc{-3.75} & \cc{-4.65} & --- & --- & --- & --- & --- & --- & ---\\
$\gamma_\mathrm{e}$ & \cc{-2.83} & \cc{-3.04} & \cc{-3.68} & \cc{-3.14} & \cc{-3.09} & \cc{-2.91} & \cc{-4.26} & \cc{-3.46} & \cc{-2.90} & \cc{-2.73} & \cc{-3.16} & \cc{-2.97} & \cc{-3.06}\\
$k_\mathrm{a}$ & \cc{-0.77} & \cc{-1.17} & \cc{-2.20} & \cc{-1.34} & \cc{-1.04} & \cc{-2.25} & \cc{-3.00} & \cc{-1.36} & \cc{-1.45} & \cc{-0.88} & \cc{-2.37} & \cc{-0.41} & \cc{-1.35}\\
$k_\mathrm{d}$ & \cc{-1.83} & \cc{-1.96} & \cc{-2.54} & \cc{-2.10} & \cc{-2.03} & \cc{-1.92} & \cc{-3.91} & \cc{-2.43} & \cc{-1.66} & \cc{-1.55} & \cc{-3.16} & \cc{-1.99} & \cc{-2.01}\\
$R_\mathrm{e}$ & \cc{-2.39} & \cc{-1.31} & \cc{-1.30} & \cc{-1.42} & \cc{-1.89} & \cc{-0.98} & \cc{-2.85} & \cc{-1.70} & \cc{-1.88} & \cc{-1.31} & \cc{-2.74} & \cc{-1.57} & \cc{-1.63}\\
$R_T$ & \cc{-1.47} & \cc{-1.91} & \cc{-2.98} & \cc{-2.07} & \cc{-1.81} & \cc{-1.93} & \cc{-2.19} & \cc{-1.88} & \cc{-1.78} & \cc{-1.54} & \cc{-2.51} & \cc{-1.34} & \cc{-1.90}\\
$\lambda_\mathrm{v}$ & \cc{-0.10} & \cc{-0.20} & \cc{-0.51} & \cc{-0.20} & \cc{-0.15} & \cc{-0.62} & \cc{-2.00} & \cc{-0.19} & \cc{-0.13} & \cc{0.14} & \cc{-1.20} & \cc{0.16} & \cc{-0.19}\\
$\epsilon_\mathrm{i}^\mathrm{CL}$ & \cc{-0.80} & \cc{-0.89} & \cc{-1.20} & \cc{-0.92} & \cc{-0.89} & \cc{-1.18} & \cc{-2.73} & \cc{-1.59} & \cc{-1.22} & \cc{-1.05} & \cc{-2.13} & \cc{-0.80} & \cc{-1.12}\\
$L_0^\mathrm{GDL}$ & \cc{-0.92} & \cc{-1.03} & \cc{-1.50} & \cc{-1.10} & \cc{-1.04} & \cc{-1.06} & \cc{-2.26} & \cc{-1.31} & \cc{-1.22} & \cc{-0.90} & \cc{-1.31} & \cc{-0.48} & \cc{-1.08}\\
$L_0^\mathrm{CL}$ & \cc{-0.81} & \cc{-1.09} & \cc{-0.42} & \cc{-0.93} & \cc{-0.97} & \cc{-1.18} & \cc{-2.71} & \cc{-1.19} & \cc{-1.11} & \cc{-0.55} & \cc{-1.77} & \cc{-0.71} & \cc{-1.03}\\
$L_0^\mathrm{PEM}$ & \cc{-0.29} & \cc{-0.41} & \cc{-0.86} & \cc{-0.47} & \cc{-0.39} & \cc{-0.83} & \cc{-2.42} & \cc{-1.19} & \cc{-1.26} & \cc{0.01} & \cc{-1.80} & \cc{-0.69} & \cc{-0.76}\\
$P_\mathrm{cl}$ & \cc{-0.88} & \cc{-0.88} & \cc{-0.98} & \cc{-0.93} & \cc{-0.92} & \cc{-0.79} & \cc{-2.31} & \cc{-1.45} & \cc{-1.24} & \cc{-0.95} & \cc{-1.41} & \cc{-0.90} & \cc{-0.94}\\
\midrule
median & \cc{-0.96} & \cc{-1.08} & \cc{-1.40} & \cc{-1.06} & \cc{-1.05} & \cc{-1.19} & \cc{-2.68} & \cc{-1.40} & \cc{-1.25} & \cc{-0.92} & \cc{-1.99} & \cc{-0.85} & \cc{-1.16}\\
	\bottomrule
	\end{tabular}}
\end{table*}

\begin{figure}
	\centering
	\includegraphics{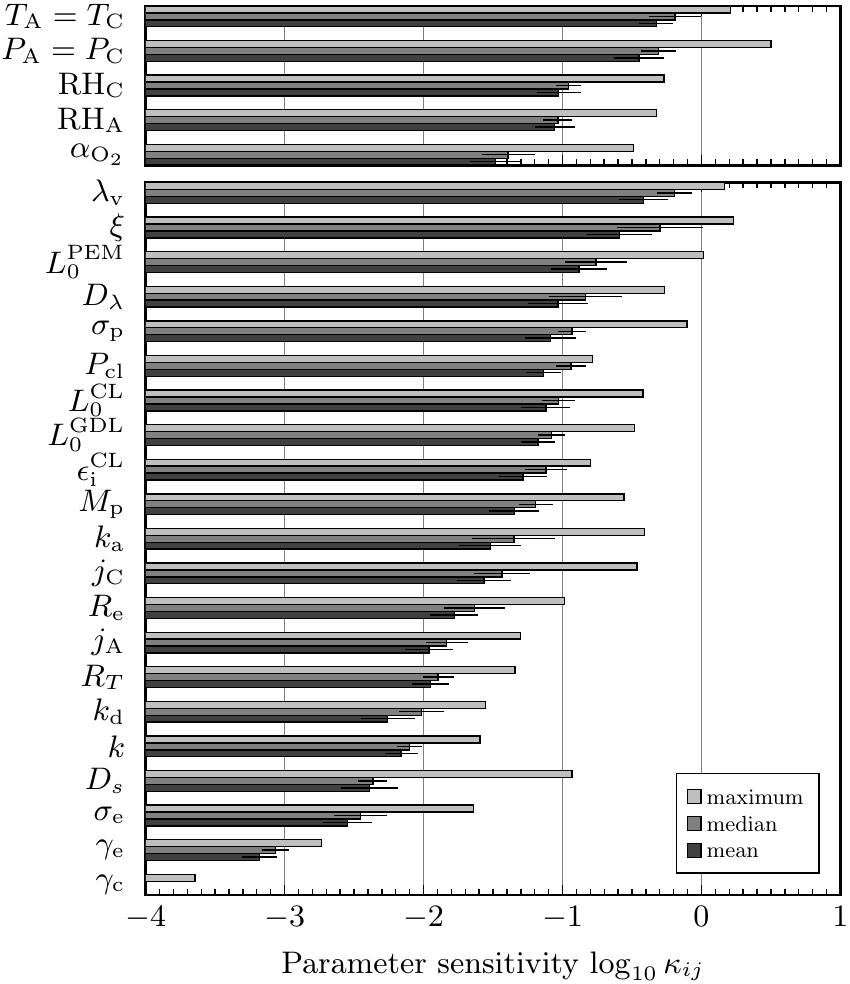}
	\caption{Global model sensitivity to parameter uncertainty. Operating conditions (top group) and MEA parameterizations (bottom group) are separately sorted by significance (i.e., by median of $\log_{10}\kappa_{ij}$ taken over all outputs $i$) in decreasing order. Error bars are standard errors of the statistics over $i$.}
	\label{fig:sensitivity_global}
\end{figure}

\begin{figure*}[tbh]
       \centering
       \includegraphics{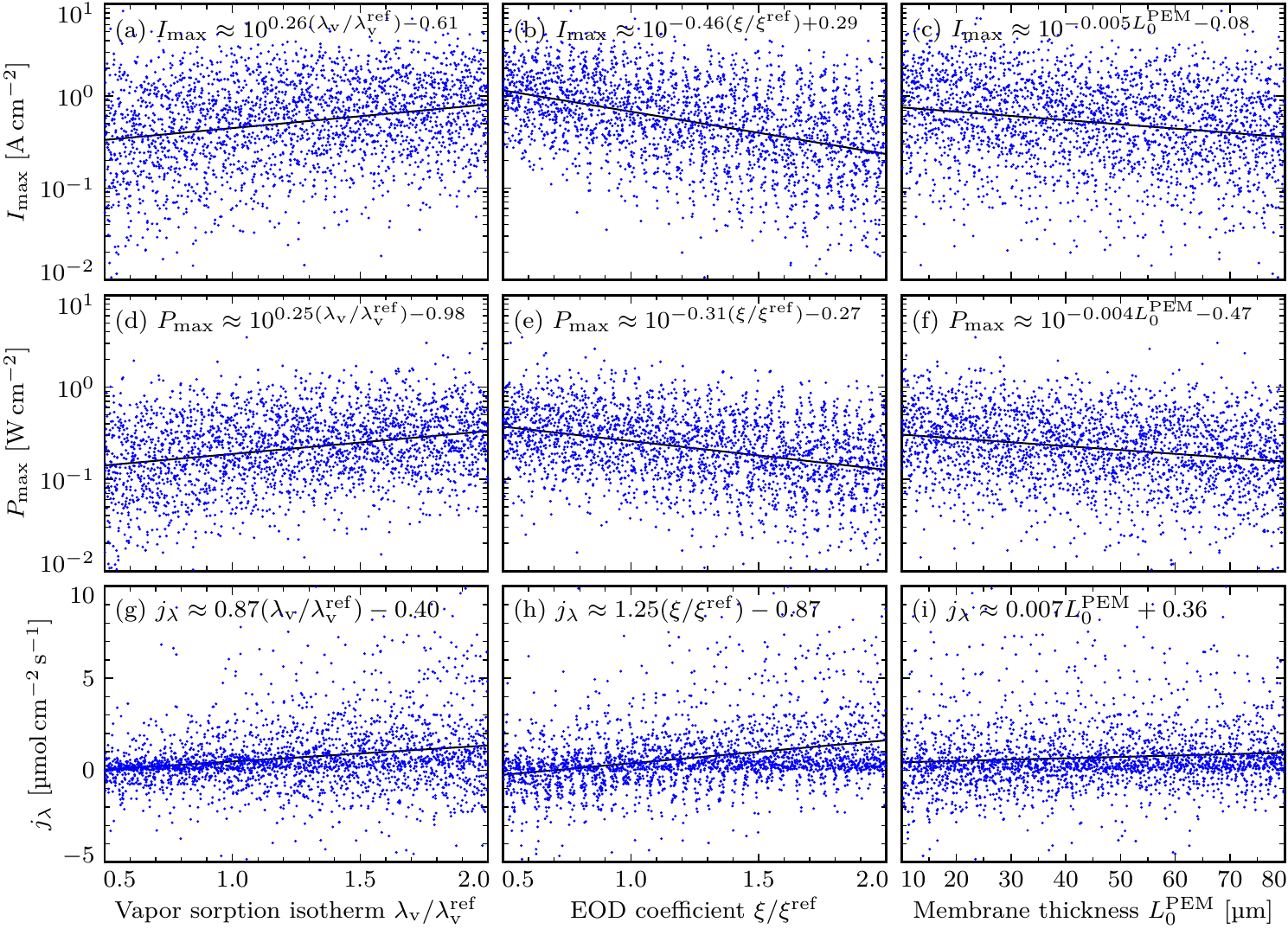}
       \caption{Scatter plots showing a selection of model responses to variation in the three parameters with highest global sensitivity. Each data point represents a sample from the Sobol sequence in the parameter ranges listed in Tab.~\ref{tab:test_intervals}. The general trend it shown with linear least-square fits in the respective lin/log spaces (solid lines) and expressed as a functional relationship in each subplot label, with all fit values in units of the shown axes.}
       \label{fig:scatter}
\end{figure*}

For the above-mentioned reasons, we seek to estimate the model sensitivity to uncertainty in all $n=26$ major parameters in a global sense in the entire $n$-dimensional domain of physically plausible values. Based on our findings of the (semi-)local uncertainty analysis shown in Figs.~\ref{fig:local_cond_opcond} and \ref{fig:local_cond_param}, and considering the ranges across which experimental material data are scattered \cite{vetter:p1}, the intervals listed in Tab.~\ref{tab:test_intervals} were chosen for the global sensitivity analysis. Most intervals are rather generous so as to not miss any critical regions. They also account for the difficulty in determining accurate estimates of the exchange current densities in Pt/C-Nafion catalysts layers \cite{neyerlin:06,neyerlin:07} and the trend toward thinner MEA layers.

In order to obtain a good estimate of global sensitivity with manageable computational effort, this $n$-dimensional parameter space needs to be sampled with high degree of uniformity -- higher than with Monte Carlo sampling. Mawardi \& Pitchumani \cite{mawardi:06} achieved this with latin hypercube sampling (LHS) in their sensitivity analysis. Here, we use the quasi-random Sobol sequence \cite{sobol:67}, which offers superior uniformity compared to LHS and the possibility to sequentially add more samples.

A Sobol sequence of length $N=2500$ is generated and then rescaled according to Tab.~\ref{tab:test_intervals}. With this set of input parameters, the MEA model is evaluated $N(n+1)$ times to get the relative condition numbers at each sample point with respect to each parameter using Eq.~\ref{eq:rel_cond_approx}. Each evaluation yields $m=12$ output values, totaling in $N(n+1)m=810{,}000$ model responses that are used for the subsequent statistical analysis. Those that are not attainable under the sampled input parameters are excluded from the analysis. For instance, small ionic conductivity of the membrane and dry gas feeds may in combination prevent the current density from reaching $1\,\mathrm{A\,cm}^{-2}$, making $U_1$ undefined, in which case only the remaining eleven key figures are used in the statistical analysis for this sample.

Tab.~\ref{tab:sensitivity_global} presents the medians of $\log_{10}\kappa_{ij}$ taken over the Sobol sequence (i.e., over all $N$ samples). Analogously to Tab.~\ref{tab:sensitivity_local}, condition numbers which are numerically indistinguishable from zero within the used error tolerances are omitted. This is the case for the condensation rate $\gamma_\mathrm{c}$, which has generally the lowest impact. Compared to the local sensitivity analysis, the overall picture changes slightly, but some specific observations remain valid. $R_\mathrm{p}$ and $j_\lambda$ are the most sensitive features, whereas $T_\mathrm{max}$ and $s_\mathrm{max}$ are the least sensitive ones. The global sensitivity of polarization curve properties generally increases toward higher current densities (lower cell voltages). Bear in mind that this is found here for the \emph{relative} condition number, which puts the derivative of $f_i$ in relation to the function value $f_i$ itself, making this a nontrivial result.

A statistical summary of the global analysis is presented in Fig.~\ref{fig:sensitivity_global}, in which both parameter categories are sorted by median over all outputs $i$ to obtain a quantitative ranking of model parameters by impact on PEMFC modeling. First in the list of operating conditions is the boundary temperature, with the gas pressure close behind. Unlike in the local analysis at reference conditions, the global race between anode and cathode gas channel relative humidities is almost tied. In the category of MEA parameterizations, the hydration isotherm of the membrane ends up first in the list with a condition number (median over all outputs $i$) of almost unity, closely followed by the EOD coefficient. On ranks three to five are the membrane thickness, water diffusivity and protonic conductivity of the ionomer. The latter is, however, only little more influential than the applied clamping pressure on rank six. Contrarily to what is sometimes commonly presumed, we find the exchange current densities only in the midrange of the parameter list, outrun by kinetic properties that have generally received less attention, such as the the vapor absorption coefficient at the ionomer--gas interface, $k_\mathrm{a}$. Just like in the local analysis, the phase change rates $\gamma_\mathrm{c}$, $\gamma_\mathrm{e}$ turn out to be the least influential in the examined range.

Although the condition number is the appropriate quantity to measure error propagation through a model, it provides no directly accessible insight into the correlation between input and output in a global sense -- such information can be extracted from scatter plots. A selection of scatter plots from our global sensitivity analysis is shown in Fig.~\ref{fig:scatter} for the three most influential MEA parameterizations ($\lambda_\mathrm{v}$, $\xi$, $L_0^\mathrm{PEM}$) and three of the most sensitive output quantities ($I_\mathrm{max}$, $P_\mathrm{max}$, $j_\lambda$). As can be recognized from Fig.~\ref{fig:scatter}a--f, the limiting current density and the peak power density increase exponentially with the membrane hydration, but they decrease exponentially with the EOD coefficient and also weakly with the membrane thickness. On the net water flux, on the other hand, the effect is more linear. Better ionomer hydration, stronger EOD and thicker membranes tend to let the water flow more from anode to cathode.

\section{Conclusion}

In Part II of this paper series, we have carried out the first full-fledged local and global parameter sensitivity analyses of a state-of-the-art two-phase PEMFC model with spatial resolution. Our work showcases the strengths of modeling: A sensitivity study of this scale, with hundreds of thousands of measurements, would be absolutely unfeasible without a numerical model. Unlike previous efforts in this direction, our study rests on a statistical analysis of a dozen model output quantities, including heat and water balance characteristics, and more than two dozen input parameters. This allowed us to compare the modeling uncertainties associated with each major model parameter in a robust quantitative way. Rather than resorting to variance-based sensitivity estimates such as Laoun et al.~\cite{laoun:16}, we have measured the propagation of uncertainty through the model explicitly by introducing the concept of condition numbers to fuel cell modeling.

Four out of the five most critical model parameters (other than the environmental conditions at which the fuel cell is operated) are constitutive transport properties of the electrolyte membrane. Considering that these are precisely those for which the experimental data available in the open literature are scattered most, as shown in Part I, this is an unfortunate circumstance that calls for better experimental characterization of the ionomer. For PEM fuel cell models to make the final leap to predictiveness, the interplay between the different water transport mechanisms and ionic conductivity must be known with high confidence.

This article answers the previously unaddressed question how much error to expect from a PEMFC model given a certain error in its constitutive material properties, both locally at given operating conditions with common MEA materials, and globally for the case that the model is employed to predict the fuel cell behavior under unexplored conditions or material substitution. While our \emph{local} analysis might be most meaningful for a typical automotive application, the results of the \emph{global} analysis are of more fundamental interest for fuel cell modelers, because a model will unfold its full potential only when being applicable to an entire range of conditions.

\section*{Acknowledgements}

Funding: This work was supported by the Swiss National Science Foundation [project no.\ 153790, grant no.\ 407040\_153790]; the Swiss Commission for Technology and Innovation [contract no.\ KTI.2014.0115]; the Swiss Federal Office of Energy; and through the Swiss Competence Center for Energy Research (SCCER Mobility).

\end{document}